\documentstyle[prl,aps,epsf]{revtex}
\begin{document}
\draft
\twocolumn
\title{Continuous loading of a magnetic trap}
\author{J. Stuhler \cite{Pfau}, P. O. Schmidt \cite{Pfau}, S. Hensler \cite{Pfau}, J. Werner \cite{Pfau}, J. Mlynek, and T. Pfau \cite{Pfau}}
\address{Fachbereich f\"ur Physik, Universit\"at Konstanz,
  Universit\"atsstra\symbol{25}e 10, D-78457 Konstanz, Germany}
\date{\today}
\maketitle
\begin{abstract}
We have realized a scheme for continuous loading of a magnetic trap
(MT). $^{52}$Cr atoms are continuously captured and cooled in a
magneto-optical trap (MOT). Optical pumping to a metastable state decouples
atoms from the cooling light. Due to their high magnetic moment (6 $\mu_B$), low-field seeking metastable atoms
are trapped in the magnetic quadrupole field provided by the MOT. Limited by
inelastic collisions between atoms in the MOT and in the
MT, we load $10^8$ metastable atoms at a rate of $10^8\,$atoms/s below 100 $\mu$K into the
MT. Optical repumping after the loading allows us to realize a MT of ground
state chromium atoms.
\end{abstract}
\pacs{32.80.Pj, 34.50.Rk}
Since their first realization \cite{Migdall:85}, magnetic traps for
neutral atoms have become important and powerful tools for many experiments in atom and quantum optics. 
Especially striking experimental results \cite{Inguscio:99} have been achieved
with Bose-Einstein condensates (BECs) that were realized by evaporatively cooling an
atomic gas in a magnetic trap (MT) \cite{Anderson:95a,Bradley:95a,Davis:95a}.
Up to the present, several groups have demonstrated pulsed \cite{Mewes:97,Martin:99}, quasi-continuous \cite{Hagley:99} or
continuous (cw) \cite{Bloch:99} outcoupling of magnetically trapped BECs.
Although multiply loading of a MT has been achieved \cite{Cornell:91}
experiments so far suffer from the absence of a method for
efficient cw loading of atoms into a BEC. Hence to date a matter wave analogon
to the continuous wave optical laser has not been realized.
Alternatively, a cw atom laser based on magnetic guiding in combination with atomic collisions was suggested \cite{Mandonnet:00a}.
In addition, cw loading of low-dimensional optical traps with laser cooled
atoms was proposed \cite{Pfau:97} and recently demonstrated \cite{Schneble:00}.

In this letter we report on the cw loading of a three dimensional 
conservative trap with laser cooled atoms that are decoupled from 
all light fields present.
We show that atoms can be optically pumped within a chromium magneto-optical
trap (MOT) \cite{Bell:99,Bradley:00a} into metastable ``dark'' states and
stored in a MT built up by the quadrupole magnetic field of the MOT. 
We present results of systematic studies on the loading
process and on the lifetime of the MT. 
Using the cw loading mechanism and a final repumping process we obtain
good starting conditions for experiments towards degenerate quantum gases with
ground state chromium atoms.

\begin{figure}[htb]
\centering
\epsfxsize0.65\columnwidth \epsfbox{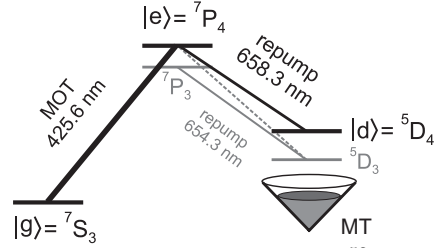}
\caption{Relevant part of the $^{52}$Cr level scheme. The MOT involves all levels and transitions, the continuous loading
  process of the magnetic trap (MT) relies on the $\Lambda$-system depicted in black (levels $|\rm g\rangle,\:|e\rangle,\:|d\rangle$).}
\label{levelscheme}
\end{figure}

Our cw loading scheme consists of an atomic reservoir and a conservative trap
overlapped in space and time.
The reservoir is prepared by cooling atoms in a MOT on a transition 
$|\rm g\rangle \rightarrow |e\rangle$ (FIG.\ \ref{levelscheme}). 
A weak decay channel
$|\rm e\rangle\rightarrow|d\rangle$ allows the transfer of reservoir
atoms into an additional long lived and trapped state $|\rm d\rangle$ in which
atoms can be accumulated.
In our realization low field seeking Zeeman substates of $|\rm d\rangle$ 
are trapped in the magnetic quadrupole field of the MOT. 
The loading can be very efficient if $|\rm d\rangle$ atoms
are decoupled from the MOT light and if their kinetic energy is smaller than
the conservative trap depth. 
A large decay rate branching ratio
($\Gamma_{\rm eg}/\Gamma_{\rm ed}\gg1$) assures a steady state MOT in thermal equilibrium and is expected to greatly reduce
reabsorption of transfer photons by atoms in the MT  \cite{Santos:00a}.
Chromium combines the desired $\Lambda$-like level scheme 
(FIG.\ \ref{levelscheme}, black levels and transitions) with a high magnetic
moment of up to 6~$\mu_B$ ($\mu_B=$Bohr's magneton). Due to its isotopic
composition (3 bosons: $^{52}$Cr (84\%), $^{50}$Cr (4\%),
$^{54}$Cr (2\%), and  one fermion: $^{53}$Cr (10\%) ) 
it is a promising element for experiments
with degenerate atomic Bose and Fermi gases. 
The magnetic dipole-dipole interaction is much stronger than for alkalis
and may lead to
to a BCS-like transition
in a degenerate Fermi gas of $^{53}$Cr \cite{Houbiers:99a}. 
In addition, chromium has technological potential in nanostructure
fabrication \cite{Drodofsky:97a,McClelland:93a} and structured doping \cite{Schulze:00} by atom lithography.
 
Magneto-optical trapping of chromium is performed on the
$^7$S$_3\rightarrow  ^7$P$_4$ transition (vacuum wavelength
$\lambda_{\rm PS} = 425.6\,$nm, decay rate $\Gamma_{\rm PS} = 31.5\times
10^6$~s$^{-1}$, saturation intensity $I_s = 8.5$~mW/cm$^2$, FIG.\
\ref{levelscheme}). Two intercombination lines connect the 
excited $^7$P$_4$ state to the metastable
states $^5$D$_4$ ($\lambda_{\rm PD4} = 
658.3$~nm) and $^5$D$_3$ ($\lambda_{\rm PD3} = 
649.3$~nm) \cite{Bell:99}. To our knowledge, the lifetime $\tau_{\rm D}$ 
of these metastable states has not been measured to date but a lower 
limit of $\tau_{\rm D}>50$~s can be deduced from our MT decay times.  
Measurements of MOT lifetimes give decay rates
of $\Gamma_{\rm PD4} = (127\pm14)$~s$^{-1}$ and $\Gamma_{\rm PD3} = (42\pm6)$~s$^{-1}$ \cite{Stuhler:00a}.
We effectively reduce the level scheme to a $\Lambda$-system
($|\rm g\rangle= ^7$S$_3$, $|\rm e\rangle= ^7$P$_4$, 
$\rm |d\rangle= ^5$D$_4$,
$\Gamma_{\rm eg} = \Gamma_{\rm PS}$, 
$\Gamma_{\rm ed} = \Gamma_{\rm PD4}$) with a
branching ratio of 250\,000 by shining in a $^5$D$_3\rightarrow   ^7$P$_3$ repumper laser.
As long as no repumper laser is applied on the 
$|\rm d\rangle \rightarrow |e\rangle$
transition atoms are optically pumped into the 
$\rm |d,m_d\rangle$ substates
($\rm m_d=-4,\dots,4$ denotes the magnetic quantum number) of 
$\rm |d\rangle$ with a significant probability of ending in low field 
seeking ($\rm m_d>0$) states. 

Our UHV-system consists of two vertically arranged chambers
connected by a Zeeman
slower. An effusion cell operated at $T_{\rm o}\sim 1700$~K is attached to the lower chamber. 
Evacuation by an ion pump and a Ti-sublimation pump leads to residual gas
pressures around $10^{-11}$~mbar in the upper chamber where the traps are located. 
Three pairs of retroreflected 1~cm diameter laser beams build up a standard
six beam $\sigma^+/\sigma^-$-light field for the MOT. 
Two coils wrapped onto the chamber produce a quadrupole magnetic field with 
gradients up to $b=20$~G/cm. 
The 4~mm diameter repumper laser beams pass the MOT and are retroreflected. 
We generate the laser light for the MOT and the Zeeman 
slower by frequency doubling a Ti:Sapphire laser using a LBO-crystal.
Two diode lasers systems serve for repumping
on the $^5$D$_3\rightarrow ^7$P$_3$ and
$^5$D$_4\rightarrow  ^7$P$_4$ ($\rm |d\rangle\rightarrow|e\rangle$) transitions
(FIG.\ \ref{levelscheme}).

We detect trapped metastable atoms by optically pumping them within a few ms 
back into the ground state $|\rm g\rangle$. 
Then $|\rm g\rangle$ atoms are resonantly excited with the MOT laser and
their fluorescence is imaged onto a calibrated CCD camera.
Repumping with the magnetic field on loads the MT with ground state chromium 
atoms \cite{Weinstein:98}.
Optical transfer of $|\rm d\rangle$ atoms into $|\rm g\rangle$ comes with a 
heating on the order of
the recoil temperature ($T_r\approx 1$~$\mu$K) due to photon scattering.
Optical pumping within the magnetic potential may change the mean magnetic
moment of trapped atoms and alter their temperature due to a variation in
potential energy. However, both effects can be neglected within our 
experimental resolution since the temperatures exceed $T_r$ by more than one order of magnitude and the mean magnetic moment
is much larger than its expected change \cite{Stuhler:00a}.  
In the experiments described here the $|\rm d\rangle$ MT can therefore be
mapped onto the $|\rm g\rangle$ MT by applying repumping light for a few ms.
 
We investigated the cw loading of the MT by performing the following
experiments. First we prepare a steady state MOT with both repumper lasers
on so that effectively no loading into the MT occurs. Then we switch off 
the $|\rm d\rangle$ repumper laser and start cw loading of the MT. 
After a variable time delay we detect the  
$|\rm d\rangle$ atoms in the MT. The resulting loading curves 
(number of MT atoms vs.\ loading time) are well fitted by
$N(t)=N_0[1-\exp(-t/\tau)]$
(fitting parameters: steady state atom number $N_0$, loading time 
constant $\tau$) to extract the MT loading rate $R = N_0/\tau$. 

\begin{figure}[htb]
\centering
\epsfxsize0.75\columnwidth \epsfbox{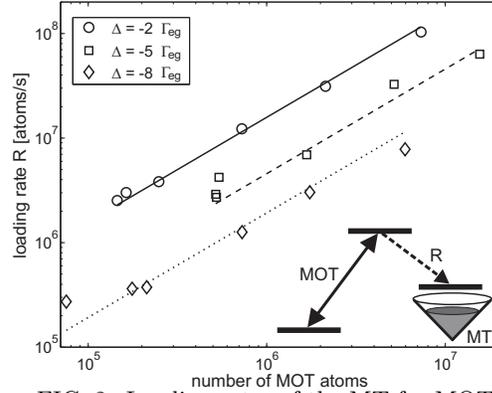}
\caption{Loading rates of the MT for MOT laser detunings of $\Delta$ =
  $-2\,\Gamma_{\rm eg}$ (circles),
  $-5\,\Gamma_{\rm eg}$ (squares) and $-8\,\Gamma_{\rm eg}$ (diamonds) as a function
  of the number of atoms in the MOT. The lines are linear least square fits to
  the data. The marker size represents the accuracy of our measurements.}
\label{loadrate}
\end{figure}

Fig.\ \ref{loadrate} shows loading rates for different
detunings ($\Delta = \omega_{\rm laser} - \omega_{\rm atom}$) of the MOT 
light and a single laser beam intensity of 15~$I_s$ vs.\ the steady state 
number of atoms in the MOT with open transfer channel. 
The number of MOT atoms was adjusted by varying $T_{\rm o}$
and/or the efficiency of the Zeeman slower. For more than a few $10^7$ MOT
atoms the collisional loss rate reaches the same order of magnitude as 
$\Gamma_{\rm ed}$. Even with all repump lasers we trap not more 
than $10^8$ atoms in the MOT \cite{Stuhler:00a}. 

For given light field parameters $R$ depends
linearly on the number of MOT atoms.
In order to evaluate the transfer efficiency $\eta=R/(N_{\rm
  MOT} P_e \Gamma_{ed})$ of the loading process, the excitation probability $P_e$ was calculated using an averaged saturation
intensity of $\langle I_s\rangle = \frac{7}{3} I_s$ as in
 \cite{Bradley:00a,Snadden:97a}. We observe $\eta = (32 \pm 5)\%, (25 \pm 4)\%$ and
$(16 \pm 4)\%$ for $\Delta=-2, -5, -8 \Gamma_{\rm eg}$.
Loading the MT at rates up to $R= 10^8$~atoms/s, we accumulate
$10^8$ atoms at peak densities of $n_0=10^{10}$~atoms/cm$^3$. 
Typical 1/e-radii of the MT are $r\sim 800$~$\mu$m while the radii of 
the Gaussian shaped MOT are about $\sigma\sim 200$~$\mu$m.  

\begin{figure}[htb]
\centering
\epsfxsize0.75\columnwidth \epsfbox{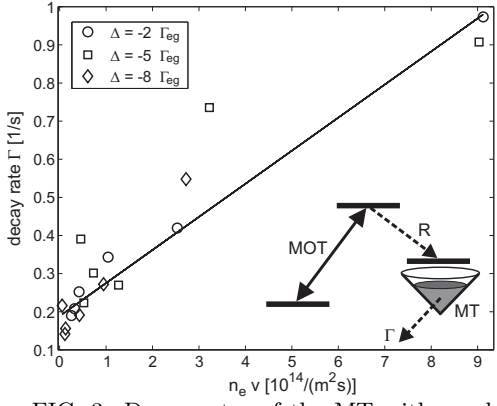}
\caption{Decay rates of the MT with overlapped MOT as a function of the effective density of excited MOT atoms n$_e$ times
  the collisional velocity $v$ for detunings of $\Delta$ =
  $-2\,\Gamma_{\rm eg}$ (circles),
  $-5\,\Gamma_{\rm eg}$ (squares) and $-8\,\Gamma_{\rm eg}$ (diamonds). The straight line is a
  linear fit to the data and gives a cross section for inelastic collisions on
  zhe order of $\sigma_{\rm ed} \sim 10^{-15}$~m$^2$.}
\label{lifetime}
\end{figure}

The maximum number of atoms in the MT is limited by the loading time 
constant of $\tau\approx1$~s observed at high loading rates. FIG.\ \ref{lifetime} shows the inverse loading time constants for the experimental parameters described above. 
The decay rates $\Gamma=1/\tau$ are corrected for ``dark'' collisions with
the residual gas and the thermal chromium beam.
This correction is done by subtracting decay rates of the 
MT in the chromium beam that were measured without MOT laser light and range 
in the shown data set from $1/20$~s to $1/2$~s depending on $T_{\rm o}$.
We plot $\Gamma$ vs.\ the product of the effective density of excited 
MOT atoms $n_e$ times the average collisional velocity $v$. 
Although measured for different detunings, $\Gamma$ increases linearly 
with $n_e v$ according to
\begin{equation}\label{gamma}
\Gamma=n_e\sigma_{\rm ed}v
\end{equation}
with a collisional cross section $\sigma_{\rm ed}$. 
This linear dependency shows that if both
traps are overlapped inelastic collisions between excited
atoms in the MOT and $|\rm d\rangle$ atoms in the MT are dominating other
loss mechanisms.
Since the MOT is much smaller than the MT
collisions occur only at the trap centre and $n_e$ can be approximated by the number of excited MOT atoms per volume of the
MT \cite{mtvolume}. The finite MOT size would give a correction factor on the
right hand side of Eq.~(\ref{gamma}) of
0.5-0.8 depending on the trap size ratio $\sigma/r$.
We assume an average collisional velocity of $v \approx [(T_{\rm MOT}+T_{\rm
  MT})\frac{8 k_B}{\pi m_{\rm  Cr}}]^{1/2}$, 
where $m_{\rm Cr}$ is the chromium mass.
We extract $\sigma_{\rm ed}\sim 10^{-15}\,$m$^2$ as an order of magnitude 
for the cross section of inelastic MOT-MT collisions by fitting the data in
  FIG.\ \ref{lifetime} linearly. This value is comparable to the
  two-body loss rate coefficient in a Cr-MOT \cite{Bradley:00a}. $\sigma_{\rm ed}$ is about one order of magnitude larger than the  values observed in mixtures of two different
  alkalis \cite{Telles:99,Schloeder:99}. Light-assisted collisions with the
  thermal chromium beam result in a non-vanishing decay rate at very low MOT densities.

\begin{figure}[htb]
\centering
\epsfxsize0.75\columnwidth \epsfbox{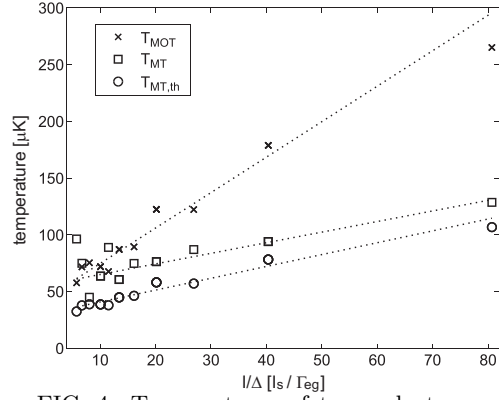}
\caption{Temperatures of trapped atoms as a function of the light shift
  parameter (six-beam intensity I/$\Delta$). T$_{\rm MOT}$ (crosses) and
  T$_{\rm MT}$ (boxes) are measured for atoms in the MOT and
  in the MT, respectively. T$_{\rm MT,th}$ (circles) are the
  theoretical temperatures of magnetically trapped atoms calculated as
  described in the text. The lines are linear least square fits to the data.}
\label{temperature}
\end{figure}

The temperature of atoms in the MT is measured in the following way.  
We pump $|\rm d\rangle$~atoms
back into the ground state $|\rm g\rangle$ with the magnetic field on. 
Then the $|\rm g\rangle$ atoms are imaged 
immediately after switching off the magnetic field.  We fit the atomic
density distribution to that of a thermal atom ensemble in a quadrupole 
magnetic field including
gravity: $n(x,y,z)=n_0\exp(-{\cal{B}}\sqrt{x^2+y^2+4z^2}-{\cal{G}}y)$. 
Here $n_0$ is the central density, ${\cal{B}}=\bar\mu b /(2 k_B T)$ and
${\cal{G}}=m_{\rm Cr} g/(k_B T)$.
The asymmetry of $n$ along $y$ (vertical axes) due to gravity is clearly
visible and the temperature is evaluated from $\cal{G}$ with an
errorbar of 10\% (fitting accuracy). 
The mean magnetic moment $\bar\mu$ (typically 
$\bar\mu = {\rm g_d \bar m_d}\mu_B=(4.5 \mbox{--}6) \mu_B$, 
corresponding to $\rm \bar m_d = 3 \mbox{--} 4$) is determined from the ratio 
$\cal{B}/\cal{G}$. The temperature of atoms
in the MT and in the MOT are plotted in FIG.\ \ref{temperature} vs.\
I/$\Delta$.
The MOT temperature, measured by
ballistic expansion of the cloud, shows the expected linear
increase with I/$\Delta$ \cite{Wallace:94}.
In the MT we observe temperatures down to 50$\,\mu$K and 
phase space densities of more than $10^{-7}$. 
MT atoms are usually colder than atoms in the MOT. 
This can be understood by using the Virial Theorem and assuming that
the transfer of MOT atoms occurs at the centre of the MT
with negligible potential energy. The initial
kinetic energy $E_i$ of MOT atoms is converted into final kinetic energy
$E_f$ and potential energy $V_f$ (for a linear potential $V_f=2 E_f$) of MT atoms:
\begin{equation}\label{virial1}
\frac{3}{2}k_B T_{\rm MOT}=E_i=E_f+V_f=3\,E_f= 3\,\frac{3}{2} k_B
T_{\rm MT,th}.
\end{equation}
Including additional initial
potential energy due to the finite size of the MOT one gets for the theoretical temperature $T_{\rm MT,th}$ of atoms in the MT:
\begin{equation}\label{t1}
T_{\rm MT,th}=\frac{1}{3} T_{\rm MOT}+ \Delta T.
\end{equation}
We estimate $\Delta T$ for a transfer from an isotropic MOT with size $\sigma$ into an
isotropic MT with mean magnetic field gradient $b$. If $\rm \bar m_d$ is the mean
magnetic quantum number of atoms in the MT, $\Delta T$ is given by 
\begin{equation}\label{t2}
\Delta T = \frac{8}{9\sqrt{2\pi}}\frac{\mu_B}{k_B} {\rm g_d \bar m_d} \, b \, \sigma
\end{equation}
and is about one order of magnitude less than $T_{\rm MOT}$.
Inserting the measured values $T_{\rm MOT}$, $\sigma$, $b$ and $\rm \bar m_d$
in Eqs. (\ref{t1}) and (\ref{t2}) we evaluate the expected 
$T_{\rm MT,th}$ (circles in FIG.\ \ref{temperature}).
Although taking our temperature resolution (about 10\%) into account, atoms
in the MT are hotter than theoretically predicted. This effect is
more pronounced at low values of I/$\Delta$ and can be explained by a
heating mechanism in the MT. Trapped $| \rm d\rangle$ atoms are heated
by 10-50 $\mu$K depending on the amount of time spent in the MT and the heating
rate described below. 

After loading, the number of atoms in the MT decays
non-exponentially indicating inelastic two-body collisions between $|\rm
d\rangle$ atoms.
In addition we observe enlargement of the trapped cloud caused by
a heating of more than 10~$\mu$K/s.
In contrast, the MT with $|\rm g\rangle$ atoms decays purely exponentially ($N(t)\propto
\exp(-t/t_0)$) with a lifetime $t_0$ of up to 60~s and shows heating rates
of only 1~$\mu$K/s. In order to distinguish between the effect of heating 
and two-body losses in the $|\rm d\rangle$ MT the standard
time derivative of the atom density is modified by a
term taking the enlargement of the cloud into account:
\begin{equation}\label{dgl}
\frac{dn_0}{dt}=-\frac{n_0}{t_0}-\beta n_0^2-\frac{n_0}{V}\frac{dV}{dt},
\end{equation} 
where $n_0$ is the peak atom density, $V$ the MT volume and $\beta$ the two-body loss rate coefficient.
We analyse our data in the following way.
The increase of the MT volume ($V=V(t)$) due to heating is fitted linearly.
After inserting this $V(t)$ and $t_0$ of the ground state MT we solve
Eq.~(\ref{dgl}) for $n_0(t)$ and fit the resulting
function (fitting parameter $\beta$) to the peak density of atoms in the
MT. This results in $\beta\sim 7\times 10^{-17}\,$m$^3$/s relatively 
insensitiv to $t_0$. The corresponding cross section $\sigma_{\rm dd}$ is
about one order of magnitude less than 
$\sigma_{\rm ed}$. Since the product of the mean MT atom
density times $\beta$ is about one order of magnitude less than the inverse 
loading time constant inelastic $|\rm d\rangle$-$|\rm d\rangle$ collisions do
not limit the number of atoms in the cw loaded MT.
To date we are not able to resolve which one of the possible mechanisms -  
magnetic field instabilities, Majorana transitions, spin flip collisions that 
release Zeeman energy or small angle collisions with atoms that leave the MT 
after a state changing collision - is dominant for the observed heating rates.
The first two processes are expected to be suppressed in a new MT with 
non-zero field minimum. 
In addition, full atomic polarization in the extreme Zeeman substate 
should lead to a reduction of $\beta$ as predicted for 
He$^*$ \cite{Shlyapnikov:94}.  

In summary, we have studied a new continuous optical loading scheme for 
conservative atom traps based on the operation of an atomic reservoir
(here a MOT) and a conservative trap (here a MT) overlapped in space and time. 
MOT atoms are transferred to the MT by a spontaneous decay into 
metastable states that are decoupled from both 
MOT light and transfer photons. 
The loading rates up to $10^8$~atoms/s depend linearly on the number
of excited MOT atoms. We continuously load up to $10^8$~atoms into the
MT, limited by collisions with excited MOT atoms. 
The lifetime of metastable atoms in the MT after switching off
the MOT is currently limited by inelastic trapped atom collisions that are 
strongly suppressed using ground state atoms.
In future experiments, the cw loading of different kinds of magnetic
traps (TOP \cite{Petrich:95} - and optical plug \cite{Davis:95a} trap) and 
the observed heating will be investigated.
Initial phase space densities on the order of $10^{-7}$ 
of $10^8$ ground state chromium atoms obtained by cw loading and
a final repumping process encourage work towards Bose-Einstein condensation.
  
This work was financed by the Deutsche Forschungsgemeinschaft and the
Optik Zentrum Konstanz. P.~O. Schmidt is supported by the
Studienstiftung des deutschen Volkes.


\begin{references}
\bibitem{mtvolume}
We define the volume $V$ of an isotropic MT with 1/e-radius $r$ by $V= 8 \pi
  r^3$.
\end{references}
\end{document}